\documentclass[12pt]{article}
\usepackage{amsmath}
\usepackage{graphicx,psfrag,epsf}
\usepackage{enumerate}
\usepackage{natbib}
\usepackage{url} 
\usepackage{xcolor}
\usepackage{bm}
\usepackage{booktabs}
\usepackage{amssymb}
\usepackage[left]{lineno}
\newcommand{\blind}{1}

	\newcommand{\innerprod}[2]{\langle #1,#2 \rangle}

\begin{document}

\def\spacingset#1{\renewcommand{\baselinestretch}%
{#1}\small\normalsize} \spacingset{1}


\if1\blind
{
  \title{\bf Incorporating non-linear mechanistic dynamics into spatio-temporal models via the SPDE approach}
  \author{L.\ Blackman$^{1,2}$\thanks{LB, PB and EJ were supported by the Scottish Government's Rural and Environment Science and Analytical Services Division (RESAS)}, J.\ Belmont$^1$, P.J.\ Bouchet$^{2*}$,J.\ Illian$^1$, E.L.\ Jones$^{2*}$,  F.\ Lindgren$^3$  \\ \small{$^1$ School of Mathematics and Statistics, University of Glasgow, UK}
     \\     
     \small{$^2$ Biomathematics and Statistics Scotland, UK}\\ 
     \small{$^3$ Department of Statistics, Lund University, Sweden}
}
  \maketitle
} \fi

\if0\blind
{
  \bigskip
  \bigskip
  \bigskip
  \begin{center}
    {\LARGE\bf Incorporating non-linear mechanistic dynamics into spatio-temporal models via the SPDE approach}
\end{center}
  \medskip
} \fi

\bigskip
\begin{abstract}
The  stochastic partial differential equation (SPDE) approach has become a popular framework for efficiently constructing spatio-temporal Gaussian random fields, yet existing formulations are largely restricted to linear equations. In this work, we develop a framework for constructing spatio-temporal models from non-linear SPDEs, enabling mechanistic dynamics to be directly incorporated into the latent structure. We build on recent advances in iterated Integrated Nested Laplace Approximations (INLA) for non-linear models, extending them to perform state and parameter estimation for two-dimensional equations and accommodate a range of observation models beyond the Gaussian case. The approach is illustrated using a spatial logistic growth model for growing and dispersing populations. Through a simulation study with point pattern data, we demonstrate that the model improves parameter recovery and predictive performance relative to linear separable and diffusion-based alternatives. We apply the model to a case study of the Eurasian Collared Dove (\emph{Streptopelia decaocto}) to demonstrate its use for invasive species. These results highlight the benefits of embedding non-linear dynamics within SPDE-based models, offering a flexible and computationally efficient framework for analysing complex spatio-temporal phenomena.
\end{abstract}

\noindent%
\vfill

\newpage
\spacingset{1.45} 
\section{Introduction}
\label{sec:intro}

Mechanistic models are widely used across the natural and physical sciences to represent processes based on established scientific principles and knowledge of the underlying mechanisms. Although these models can capture complex, non-linear dynamics, they are not primarily driven by data. In contrast, statistical models use data to infer patterns and quantify uncertainty, but they often provide limited insight into the processes generating the observed behaviour. Historically, these two modelling traditions have developed largely independently; however, increasing work has been done to combine mechanistic process understanding with data-driven inference. 

Epidemiology has been a particularly active area for such developments, with many models integrating mechanistic descriptions of disease transmission with statistical inference (see, e.g., \citealp{Lloyd-Smith-2015} and the papers therein). These models are typically non-spatial, and when spatial structure is included it is usually represented on a discrete domain. Similar ideas have emerged in ecology through dynamic species distribution models, which incorporate population growth and range expansion processes into statistical frameworks \citep{Pagel-2012, Hooten-2008}. However, these approaches also generally rely on discrete grid-based representations of space. More recently, attention has shifted towards methods that operate over continuous spatial domains. \cite{Wikle-2026} provides an overview of these emerging mechanistic-statistical approaches. Many recent methods are based on physics-informed neural networks, which can improve predictive performance but often sacrifice interpretability and may be less efficient and accurate than traditional finite-element approaches \citep{Grossmann-2024}. Consequently, there remains a need for computationally efficient and interpretable methods that integrate non-linear mechanistic models with spatio-temporal statistical frameworks across continuous space. Furthermore, the many of the approaches mentioned above are restricted by a lack of general-use software for applied users \citep{Swallow-2022}. In this paper, we address these challenges by developing a latent Gaussian SPDE framework for non-linear differential equations on two-dimensional continuous spatial domains. This can be used within \texttt{R-INLA}, a widely adopted and extensible platform for Bayesian inference\citep{Rue-2009, Lindgren-2015}.


\subsection{Spatio-temporal statistics}
 
Many real-world processes vary continuously across space and time, and spatio-temporal models allow us to understand and predict their behaviour \citep{habereder-2026}. Examples include the distribution of wildlife populations \citep[e.g.,][]{Yuan-2017}, the spread of infectious diseases \citep[e.g.,][]{Khan-2026}, and the dispersion of air pollutants \citep[e.g.,][]{Cameletti-2013}. In these models, a common approach to representing continuous spatio-temporal dependence is to use latent Gaussian random fields (GRFs), typically specified using a covariance function, either separable (space and time are independent) and non-separable (space and time are dependent). Considerable effort has been devoted to the development of flexible families of spatio-temporal covariance functions capable of capturing complex dependence structures, such as the non-separable methods of \cite{Gneiting-2002} and \cite{Cressie-1999}. However, specifying a covariance function directly can lead to substantial computational challenges –- particularly for large spatio-temporal datasets, where dense covariance matrices require factorising \citep{Louzada-2025, Simpson-2011}. In recent years, these challenges have motivated the development of alternative representations of Gaussian fields \citep{Lindgren-2022}. The Stochastic Partial Differential Equation (SPDE) approach \citep{Lindgren-2011, Lindgren-2024-SPDE} offers one method for generating GRFs by representing Gaussian fields as solutions to differential equations using the finite element method, leading to sparse precision matrices and substantial computational gains. This framework has been widely adopted due to its computational efficiency and inclusion in the \texttt{R-INLA} package \citep{Rue-2009, Lindgren-2015}. Recent work has begun to integrate mechanistic dynamics within the SPDE framework. \cite{Mandujano-Reyes-2024} showed that the diffusion extension of the Matérn equation is equivalent to an ecological diffusion equation, and could therefore be used as a model of disease spread, and \cite{Clarotto-2024} used an advection-diffusion equation to model the mechanisms underpinning solar radiation data. Both these approaches created mechanism-informed geostatistical models that retains the computational efficiency of the sparse Gaussian Markov random field representation while incorporating knowledge of the underlying physical process. 

However, existing SPDE-based approaches are limited to linear equations but many physical and biological systems are governed by non-linear processes. Incorporating such non-linear dynamics within the SPDE framework would extend its applicability and improve predictive power while retaining its computational advantages. Although recent work has demonstrated how non-linear differential equations can be incorporated into latent Gaussian models using iterated Integrated Nested Laplace Approximation (INLA), the methodology has so far been restricted to one-dimensional systems with Gaussian observations \citep{Anderka-2024}.  Many practical applications involve processes evolving over two-dimensional spatial domains with more complex observation processes (e.g.\ modelling the distribution of a species with semi-continuous data \citep{Jones-Todd-2017}; modelling the spread of an infectious disease with active and passive surveillance \citep{Nelli-2020}), motivating the development of more generalised methods.

\subsection{Contribution}
In this work, we develop, for the first time, a two-dimensional extension of the latent Gaussian modelling framework for non-linear differential equations, allowing spatio-temporal random fields to be constructed directly from mechanistic descriptions of the underlying data-generating process. This allows scientifically meaningful parameters, such as growth or dispersal rates, to be estimated directly from data while accounting for uncertainty, and produces predictions that are constrained by the scientific principles governing the system. 

We demonstrate the proposed framework with a specific example of the spatial logistic growth equation, one of the most widely used formulations in population dynamics and mathematical biology \citep{Verhulst-1844, Schurz-2007, Karim-2023}. Our primary motivation for this example is to model increasing and dispersing wildlife populations, particularly invasive species colonising new areas, where local population growth and dispersal jointly drive colonisation of new areas. The logistic growth equation provides a natural framework for these dynamics because it incorporates both population expansion and density-dependent regulation through a local carrying capacity, a key feature of realistic invasion processes \citep{Morris-2013, Jones-2020}. Unlike conventional covariance-based approaches, incorporating local population growth directly into the model allows biological knowledge about reproduction and resource limitation to influence the spatio-temporal dependence structure. Although our focus is on invasive species, the logistic growth model is also applicable to a wide range of spatio-temporal phenomena, including infectious disease dynamics \citep{Cramer-2004, Shen-2020} and wildfires \citep{Podschwit-2018}.

 In section~\ref{sec:sim-study}, we use simulated data to demonstrate how the proposed model outperforms existing separable and non-separable spatial-temporal covariance models in parameter estimation and state prediction. In section~\ref{sec:dove}, we apply the logistic growth model to a case study of the Eurasian Collared Dove (\emph{Streptopelia decaocto}), an invasive species in North America. Alongside the methodological developments, we introduce the R package \texttt{INLAloggrowth}, which implements the proposed logistic growth model and makes it readily accessible for ecological applications. Built on \texttt{inlabru} \citep{Lindgren-2024-inlabru} and \texttt{R-INLA} \citep{Rue-2009}, it offers computationally efficient inference and flexible specification of observation models, making it well suited to a wide range of  applications. The latent models implemented in \texttt{INLAloggrowth} can be used with the broad range of likelihood functions available in \texttt{R-INLA}, allowing alternative observation models to be explored with minimal changes to the code. The \texttt{inlabru} package allows for straightforward construction of joint models for different data sources and for non-linear predictors in the observation process. This is particularly valuable for ecological data, such as distance sampling \citep{Yuan-2017} and citizen science data \citep[e.g.,][]{Panunzi-2024, Pacifici-2017}.

\section{Methods}
\label{sec:meth}
\subsection{The SPDE approach for linear models}
The SPDE approach proposed by \cite{Lindgren-2011} generates a Gaussian field from a linear differential equation. It derives a sparse precision matrix from a 2-dimensional SPDE of the form: 
\begin{align}\label{eq:SPDE-generic}
    Df(\bm{s}) = W(\bm{s}) \qquad \bm{s} \in D \subset \mathbb{R}^2
\end{align}
describing the spatial covariance structure of an unknown stochastic process $f(s)$ in terms of a linear differential operator $D$ and Gaussian white noise $W(\bm{s})$, which can be correlated or uncorrelated in space. Most applications use the Matérn equation as it provides a flexible description of smooth spatial variation \citep{Porcu-2024}. A growing body of research has further extended the Matérn model to account for increasingly realistic spatial structure, for example the barrier model of \cite{Bakka-2019} and the non-stationary models of \cite{Ingebrigtsen-2014} and \cite{Mandujano-Reyes-2024}.

Using inner product notation $\innerprod{f}{g} = \int f(\bm{s})g(\bm{s})d\bm{s}$ where the integral is over the spatial domain, the weak solution to the SPDE \eqref{eq:SPDE-generic} is a random field $f$ that satisfies
\begin{align}
    \innerprod{Df}{\psi} \overset{d}{=} \innerprod{W}{\psi}
\end{align}

where $\overset{d}{=}$ means ``converges in distribution to'' (following the notation of \cite{Lindgren-2011}) and $\psi$ is any smooth function with compact support in $\mathbb{R}^2$, known as a test function. An approximate weak solution is found using finite element methods. This uses a graph (also referred to as a ``mesh" in the literature) to split the domain into elements and a basis function $\phi_j$ is associated with each point $j;  j =1, ... ,n_s$ on the mesh. The basis functions are typically piecewise-linear functions defined such that $\phi_j$ is $1$ at point $j$, decreasing linearly to 0 along the neighbouring edges. Then, by approximating both the random field $f$ and the test function $\psi$ as linear combinations of the basis functions:

\begin{align}\label{eq:fem-approx}
    \sum_{i=1}^{n_s}\sum_{j=1}^{n_s} u_i\innerprod{D\phi_i}{\phi_j} &\overset{d}{=}  \sum_{i=1}^{n_s}\sum_{j=1}^{n_s} W_i\innerprod{\phi_i}{\phi_j}\\
    \bm{Du} &= \bm{W},
\end{align}

where $\bm{D}$ has $(i,j)^{th}$ entry $\innerprod{D\phi_i}{\phi_j}$, $\bm{u}$ is the vector of random weights $(u_1,...,u_{n_s})^T$ and $\bm{W}$ is a vector drawn from the spatial noise $(W_1,...,W_{n_s})^T$. It follows that:
\begin{equation}\label{eq:SPDE-prior}
    \bm{u} \sim N(\bm{0, Q_u^{-1}}),
\end{equation} where $\bm{Q_u} = \bm{D^TQ_e D}$ and $\bm{Q}_e$ is the precision matrix for $W$. Hence the SPDE approach is a way to place a multivariate Gaussian prior on $\bm{u}$, where the precision matrix $\bm{Q}_u$ encodes the covariance structure described by the differential operator $D$ and noise process $W$. $\bm{Q}_u$ can be approximated by a sparse matrix, making this approach computationally efficient particularly for INLA-based inference.

\subsection{INLA}\label{sec:INLA}
INLA is a framework for fast approximate Bayesian inference for Latent Gaussian Models (LGMs) \citep{Rue-2009}. An LGM consists of observations $\bm{y}$, a latent Gaussian vector $\bm{\phi}$ and hyperparameters $\bm{\theta}$, and can be expressed hierarchically as:

\begin{align}
    \bm{y| \phi, \theta} &\sim p(\bm{y|\phi,\theta}) \\ \nonumber
    \bm{\phi|\theta} &\sim N(\bm{\mu_\phi}, \bm{Q(\theta)}^{-1}) \\ \nonumber
    \bm{\theta} &\sim p(\bm{\theta}).
\end{align}

The latent field $\bm{\phi}$ may contain a Gaussian Markov random field (GMRF) generated using the SPDE approach, together with covariate effects and other model components. INLA performs Bayesian inference by constructing a sequence of Laplace and Gaussian approximations to the posterior distributions of the latent field and hyperparameters. The joint posterior density of the hyperparameters is calculated by 
\begin{equation}
    \hat{p}(\bm{\theta|y}) \approx \left. \frac{p(\bm{\theta, \phi, y})}{\tilde{p}_G(\bm{\phi|\theta, y})} \right |_{\phi = \phi^*(\theta)},
\end{equation}
where $\tilde{p}_G(\bm{\phi|\theta, y})$ is the Gaussian approximation of $p_G(\bm{\phi|\theta, y})$ about the mode $\bm{\phi^*(\theta)}$ for a given $\bm{\theta}$. 

The marginal posterior distribution of a latent component $\phi_i$ is then obtained by numerical integration over the hyperparameters,
\begin{equation}
\hat{p}(\phi_i | \bm{y}) = \sum_k \hat{p}(\phi_i | \bm{\theta}_k, \bm{y}) \hat{p}(\bm{\theta}_l | \bm{y}) \Delta_k,
\end{equation}
where $\bm{\theta}_k^{(n)}$ are the posterior marginal integration nodes and $\Delta_k$ are the corresponding integration weights. The approximate density $\hat{p}(\phi_i | \bm{\theta}, \bm{y})$ is calculated using another Laplace approximation:
\begin{equation}
    \hat{p}(\phi_i|\bm{\theta,y}) \propto \left. \frac{p(\bm{\phi, \theta, y}}{\tilde{p}_G(\bm{\phi_{-i}}|\phi_i, \bm{\theta,y})} \right |_{\bm{\phi_{-i}} = \bm{\phi_{-i}^*}(\phi_i,\bm{\theta})}
\end{equation}
where $\tilde{p}_G(\bm{\phi_{-i}}|\phi_i, \bm{\theta,y})$ is the Gaussian approximation to $p(\bm{\phi_{-i}}|\phi_i, \bm{\theta,y})$ about the mode $\bm{\phi_{-i}^*}(\phi_i,\bm{\theta})$. 
The SPDE approach is particularly well suited to INLA because it yields sparse precision matrices $\bm{Q}(\bm{\theta})$. Exploiting this sparsity substantially reduces the computational cost of the Gaussian approximations and numerical integration steps, enabling efficient inference for large spatio-temporal models.

\subsection{Iterated INLA for State and Parameter Estimation in non-linear Dynamical Systems} \label{Changing-SPDE}
Anderka, Deisenroth and Takao \citeyearpar{Anderka-2024} propose a method for estimating the state and parameters of a one-dimensional non-linear spatio-temporal SPDE from Gaussian observations using INLA. The equation is linearised using a Taylor expansion about an unknown linearisation point in order to allow discretisation using finite differences in space and time. This approximation represents the non-linear dynamics as locally linear but retains the global non-linearity, and defines a Gaussian prior for the one-dimensional random field given the linearisation point \citep[see section 3.2 of][]{Anderka-2024}. The model is then iteratively fit to estimate the linearisation point, with a choice of two update rules, discussed in Section \ref{sec:2d-extension}. 

This is a similar idea to the approach implemented in \texttt{inlabru} \citep{Lindgren-2024-inlabru}, which also fits non-linear models by linearising using a Taylor expansion and iterative fitting. However, \texttt{inlabru} takes non-linear transformations of the model components and hence the non-linearity only appears in the observation model. 
In contrast, the Iterated INLA method constructs priors for the latent process, allowing non-linear differential equations to generate the covariance structure of the random field. This allows the temporal structure to better reflect the mechanisms that generate the data, and \cite{Anderka-2024} showed this improves predictive power. Generalising this approach would extend these benefits to a wider range of two-dimensional models and applications where the underlying dynamics are inherently non-linear.

\subsection{Gaussian Markov Random Fields from 2D Non-linear SPDEs}\label{sec:2d-extension}

In this work, we expand the previous developments of \cite{Anderka-2024} to two-dimensional non-linear SPDEs, allowing the method to be more applicable to a broader range of problems. We also generalise the framework to allow for non-Gaussian observations. 

Consider a non-linear SPDE of the form:
\begin{align}\label{eq:spde-form}
    Lu(\bm{s},t) = W(\bm{s},t) \qquad \bm{s} \in \mathcal{D} \subset\mathbb{R}^2, 0 \leq t \leq T,
\end{align}
where the differential operator is $L = \frac{\partial}{\partial t} - D$. $D$ is some non-linear operator, which may depend on unknown parameters $\bm{\theta}$ to be estimated jointly with the latent field $u(\bm{s},t)$, and $W(\bm{s},t)$ is a spatio-temporal noise process. To enforce positive smoothness in space and time,  $W(\bm{s},t)$ cannot in general be white in both space and time, instead requiring some form of spatial structure to prevent generating non-smooth field.

In line with \cite{Anderka-2024}, we linearise the operator $D$ by a first order Taylor expansion about a linearisation point $u_0^{(n)}$:
\begin{align}
    D[u] &\approx D(u_0^{(n)}) + D_0^{(n)}(u-u_0^{(n)}) \nonumber \\
    &= (D-D_0^{(n)})u_0^{(n)} + D_0^{(n)}u
\end{align}

Then $L$ can be approximated by
\begin{align} \label{eq:linear}
    L[u] &\approx L_0^{(n)}u-r_0^{(n)},
\end{align}
where
\begin{align*}
    L_0^{(n)}u &= \frac{\partial u}{\partial t} - D_0^{(n)}u  \nonumber\\
    r_0^{(n)} &= D[u_0^{(n)}] - D_0^{(n)}u_0^{(n)} \nonumber \\
    &= L_0^{(n)}u_0^{(n)} - L[u_0^{(n)}] \nonumber
\end{align*}

We seek an approximate weak solution to the SPDE $L_0^{(n)}u(\bm{s},t) = r_0^{(n)}+ W(\bm{s},t)$, satisfying
\begin{align}
    \innerprod{L_0^{(n)}u(\bm{s},t)}{\psi} \overset{d}{=} \innerprod{r_0^{(n)}}{\psi} + \innerprod{W(\bm{s},t)}{\psi}
\end{align}
for some test function $\psi$. Using a finite element approximation with $M$ linear basis functions as in equation \eqref{eq:SPDE-prior}, we can discretise the field in space: 
\begin{align}
    \sum_{i=1}^{n_s}\sum_{j=1}^{n_s}\innerprod{L_0^{(n)}u_i(t)\phi_i}{\phi_j} \overset{d}{=} \sum_{i=1}^{n_s}\sum_{j=1}^{n_s} \innerprod{r_{0,i}^{(n)}(t)\phi_i}{\phi_j}+ \innerprod{W_i(t)\phi_i}{\phi_j}
\end{align}

To approximate the field in time, we use the Backwards Euler method, dividing the temporal domain into $n_t$ equally spaced points $\{T_1, T_2, ... T_{n_t}\}$, such that $T_{n+1} - T_n = h\quad \forall n < n_t$. The derivative at time $T_{n+1}$ is approximated using the current and previous time points:
\begin{equation}
    \frac{d u_i}{d t}\Bigr|_{t = T_{n+1}} \approx \frac{u_i(T_{n+1}) - u_i(T_n)}{h}.
\end{equation} Euler methods usually require a known initial condition, typically fixing one of the endpoints of the temporal domain. As each step in the Backwards Euler method looks backwards to the previous timepoint, we require some information about the initial state $\bm{u}(T_1)$. Here, we use this recursive structure to relax the assumption of a known initial state and treat the initial condition as uncertain. Rather than assume $\bm{u}(T_1)$ is fixed, we estimate the initial state from the data and supply this as a prior to propagate the uncertainty. We fit a spatial model to the observations from the first time point, then use the resulting GMRF $\bm{u}(T_1) \sim N(\bm{\mu}, \bm{P}^{-1})$ as a Gaussian prior for the initial condition. We can then propagate the uncertainty by changing the structure of the white noise process $W(\bm{s},t)$ such that it has precision $\bm{P}$ at time $T_1$.  Applying the time discretisation, we obtain a Gaussian prior for $u$ in the same way as \eqref{eq:fem-approx}, where $\bm{Q}$ is the modified precision matrix for $W$: 
\begin{align}
    \bm{L_0^{(n)}u} &=\bm{r_0^{(n)}} + \bm{W} \\ \nonumber
    \bm{u} \sim N(\bm{ L_0^{(n)^{-1}}}&\bm{r_0^{(n)}}, (\bm{L_0^{(n)\top}QL_0^{(n)}})^{-1})
\end{align}

The linearisation point is estimated by iteratively fitting the model. Since the operator $D$ may depend on unknown parameter $\bm{\theta}$, inference is performed jointly for the latent field $\bm{u}$ and the operator parameters $\bm{\theta}$. \cite{Anderka-2024} propose two update rules for the one-dimensional case with Gaussian observations $\bm{y}$, both of which we extend to the two-dimensional case with any observation process that can be adequately approximated using the Laplace Approximation. The ``Type I update rule'' is a simple weighted average of the $n^{th}$ linearisation point and the posterior mean of the latent field after the $n^{th}$ model fitting, so is independent of the observation process. Let $\bm{\theta}_k^{(n)}$ be the posterior marginal integration nodes for the differential operator parameters, $\Delta_k$ the corresponding integration volume elements, $\tilde p(\bm{\theta_k}^{(n)}|\bm{y})$ the marginal posterior probability of $\bm{\theta}_k^{(n)}$, and $\bm{\mu_{u|y}}(\bm{\theta}_k^{(n)})$ the posterior mean for the random field $\bm{u}$ at $\bm{\theta}_k^{(n)}$, as in Section~\ref{sec:INLA}. Then the Type I update rule is:
\begin{align}
    &\bar{u}^{(n)} = \sum_k \bm{\mu_{u|y}}^{(n)}(\bm{\theta}_k)\tilde p(\bm{\theta_k}^{(n)}|\bm{y})\Delta_k \\
    &\bm{u}_0^{(n+1)} = (1-\gamma)\bm{u}_0^{(n)} + \gamma \bm{\bar{u}} ^{(n)},
\end{align}
$\gamma \in (0,1]$ is a tunable damping parameter chosen somewhat arbitrarily to balance fast convergence and computational stability.

The ``Type II update rule'' is as follows:
\begin{align}
        &\bm{\bar{P}}^{(n)} = \sum_k \bm{P_{u|y}}^{(n)}(\bm{\theta}_k^{(n)})\tilde p(\bm{\theta_k}^{(n)}|\bm{y})\Delta_k \nonumber \\
        & \bar{\bm{b}}^{(n)} = \sum_k\bm{P_{u|y}}^{(n)}(\bm{\theta}_k^{(n)}) \bm{\mu_{u|y}}(\bm{\theta}_k^{(n)}) \tilde p(\bm{\theta_k}^{(n)}|\bm{y})\Delta_k \nonumber \\
        & \bar{\bm{u}}^{(n)} = (\bm{\bar{P}}^{(n)})^{-1}\bar{\bm{b}}^{(n)} \nonumber \\
        & \bm{u}_0^{(n+1)} = (1-\gamma)\bm{u}_0^{(n)} + \gamma \bm{\bar{u}} ^{(n)},
    \end{align}
where  $\tilde p(\bm{\theta_k}^{(n)}|\bm{y})$ is the marginal posterior probability of $\bm{\theta}_k^{(n)}$, and $\bm{\mu_{u|y}}(\bm{\theta}_k^{(n)})$ and $\bm{P_{u|y}}^{(n)}(\bm{\theta}_k^{(n)})$ are the mean and the precision matrix for the random field $\bm{u}$ at $\bm{\theta}_k^{(n)}$. \cite{Anderka-2024} prove this rule is an approximate Gauss-Newton method for minimising $\mathbb{E}(-\log p(\bm{u|y,\theta})$ when the observations $\bm{y}$ are Gaussian; we prove that the same holds for any observation process that can be adequately approximated using the Laplace approximation in Supplementary Material S1. Consequently, the proposed framework can be combined with any observation family implemented in \texttt{R-INLA}, substantially broadening its applicability across different data types and contexts.

\section{Application to the Logistic Growth Equation}\label{sec:example}
To illustrate the proposed methodology, we apply it to the logistic growth equation. This equation is widely used to model increasing and spreading biological populations, including invasive and introduced species \citep{Verhulst-1844, Schurz-2007, Karim-2023, Hooten-2008}. It captures several key features of spatial population dynamics, including density dependent growth, dispersal, and the concept of a local carrying capacity, defined as the maximum population size that can be sustained by the resources available in the environment.  It also provides an opportunity to demonstrate the flexibility of the framework across different observation models, as ecological data are often collected through diverse observation processes. Different observation processes may yield, for example, count data, presence-absence records or presence-only observations from the same underlying population; all can be accommodated within the INLA framework. 
The logistic growth equation therefore provides a useful example for illustrating how physically motivated non-linear SPDEs can be incorporated into spatio-temporal models.
We take the following form of the logistic growth equation as the basis for our model:
\begin{equation} \label{eq:growth-diffusion}
    \frac{\partial u}{\partial t} - \alpha u\left(1-\frac{u}{K}\right) + \nabla\cdot \xi \nabla u = \sigma uW(\bm{s},t),
\end{equation}
where $u(\bm{s},t)$ is the number of animals, $\alpha$ is the growth rate, $K$ is the carrying capacity, $\xi$ is the dispersal coefficient, and $W(\bm{s},t)$ is a noise term, which is Matérn correlated in space and white in time to give smoothness 1/2 in time and 1 in space. As mentioned in section \ref{sec:2d-extension}, the noise process cannot be completely unstructured (i.e.\ white in space and time) in this case, as this gives smoothness 0 in space and time. Whilst this is not in the form of \eqref{eq:spde-form} due to the term $\sigma uW(\bm{s},t)$  on the right hand side, it can be rearranged into this form after a log transformation. The $\sigma uW(\bm{s},t)$ term makes the stochastic variation density dependent, which is both algebraically convenient for our purposes and more biologically realistic \citep{Schurz-2007}, as larger populations typically experience greater absolute variability than smaller populations. For example, year-to-year fluctuations of tens of individuals may be common in a population of 1000 animals but are unlikely in a population of only 15 animals. Consequently, the variance of the stochastic component increases with population size rather than remaining constant across all densities.

To enforce $u>0$, consider a log transformation $u = \exp(v)$. We derive the mean and covariance for $v$, which can then be included as a random effect in the linear predictor for any observation process. Working on the log scale is useful for interpretability in cases with a log link; for example a Log Gaussian Cox Process (LGCP) could have the log intensity equal to the log number of animals. The full derivation of the GRF of the log number of animals, $\bm{v} \sim N(\bm{L^{-1}}\bm{r}_0^{(n)}, (\bm{L}^\top\bm{Q}\bm{L})^{-1})$, is given in Supplementary Material S2.

\subsection{Simulation Study}\label{sec:sim-study}

The proposed logistic growth model is implemented in \texttt{R-INLA} and is available in the R package \texttt{INLAloggrowth}. Built as a wrapper around \texttt{inlabru}, the package is fully compatible with the observation families and non-linear predictor framework available in \texttt{inlabru}. It automates the iterative fitting procedure described in Section~\ref{sec:2d-extension}, enabling efficient inference for the logistic growth model with minimal user intervention. We use the \texttt{cgeneric} framework within \texttt{R-INLA} to perform the computationally intensive calculations, such as calculating the precision matrix, using sparse matrix algebra in C for scalability and speed. All model fitting was performed on the Crop Diversity HPC \citep{Percival-Alwyn-2024}.

To compare the proposed model with existing methods, we perform a simulation study on point pattern observations. We aim to infer the state and parameters of this equation from point pattern data, motivated by the increasing use of citizen science observations in ecological spatial models \citep[see e.g.,][]{Seaton-2024, Pacifici-2017, Panunzi-2024}). In such settings, data are typically collected as individual sightings with spatial locations, rather than as aggregated counts. We simulate 10 years of point pattern observations from the spatial logistic growth model over a 1x1 square, fit the models on the first five years and test the out-of-sample forecasting abilities on the latter five. We focus only on the case where the data are generated by the logistic growth equation, as the problem of model misspecification for logistic growth-like models is an ongoing research topic \citep{Browing-2025, Simpson-2022, Simpson-2024}.

In addition to the logistic growth model, we fit 3 other models to the simulated data. We consider two separable spatio-temporal random effects: an IID Matérn model for each time point, and the Kronecker product of the Matérn model and a first order autoregressive (AR1) model in time. We also consider the non-separable critical diffusion model from \cite{Lindgren-2024-SPDE}, as it is the most similar to the proposed model. In order to obtain parameter estimates for the growth rate and carrying capacity for these models, we use the analytical solution to the non-spatial logistic growth equation: 

\begin{equation} \label{logistic-growth-solution}
  p(t) = \frac{Kp_0\exp({\alpha t})}{K-p_0 + p_0\exp({\alpha t})} \qquad t>0,
\end{equation}
where $p_0$ is the (estimated) initial population. It is not possible to directly estimate the dispersal parameter $\xi$ using a Matérn model, as Matérn models have a spatial range parameter representing the distance at which the correlation is 0.1. The dispersal parameter $\xi$ is a diffusion coefficient representing the rate at which the population spreads through space per unit of time, so is a spatio-temporal parameter rather than a purely spatial parameter. For the Matérn models, the intensity function of the LGCP is $\log ( \lambda(\bm{s}, t)) = \log(p(t)) + \zeta(\bm{s},t)$ for $t>0$, where $p(t)$ is the deterministic logistic growth solution and $\zeta(\bm{s}, t)$ is a spatio-temporal random effect, modelled using either an IID, AR1, or diffusion structure. For the logistic growth model, $\log( \lambda(\bm{s}, t)) = v(\bm{s}, t)$, where $v$ is the log number of animals. The same mesh, priors, and starting values were used for each model. Since the data are simulated, the true intensity function is known, allowing posterior predictive performance to be assessed by comparing predicted and true intensities. We calculate three strictly proper scores: the mean squared error (MSE), the median absolute error (MAE) and the continuous ranked probability score (CRPS) \citep{Matheson-1976}. Similarity between the mean predicted and true intensity maps was quantified using the Structural Similarity Index (SSIM) \citep{Jones-2016, Wang-2004}.

\subsubsection{Results}
\begin{figure}[hbt!]
    \centering
    \includegraphics[width=\linewidth]{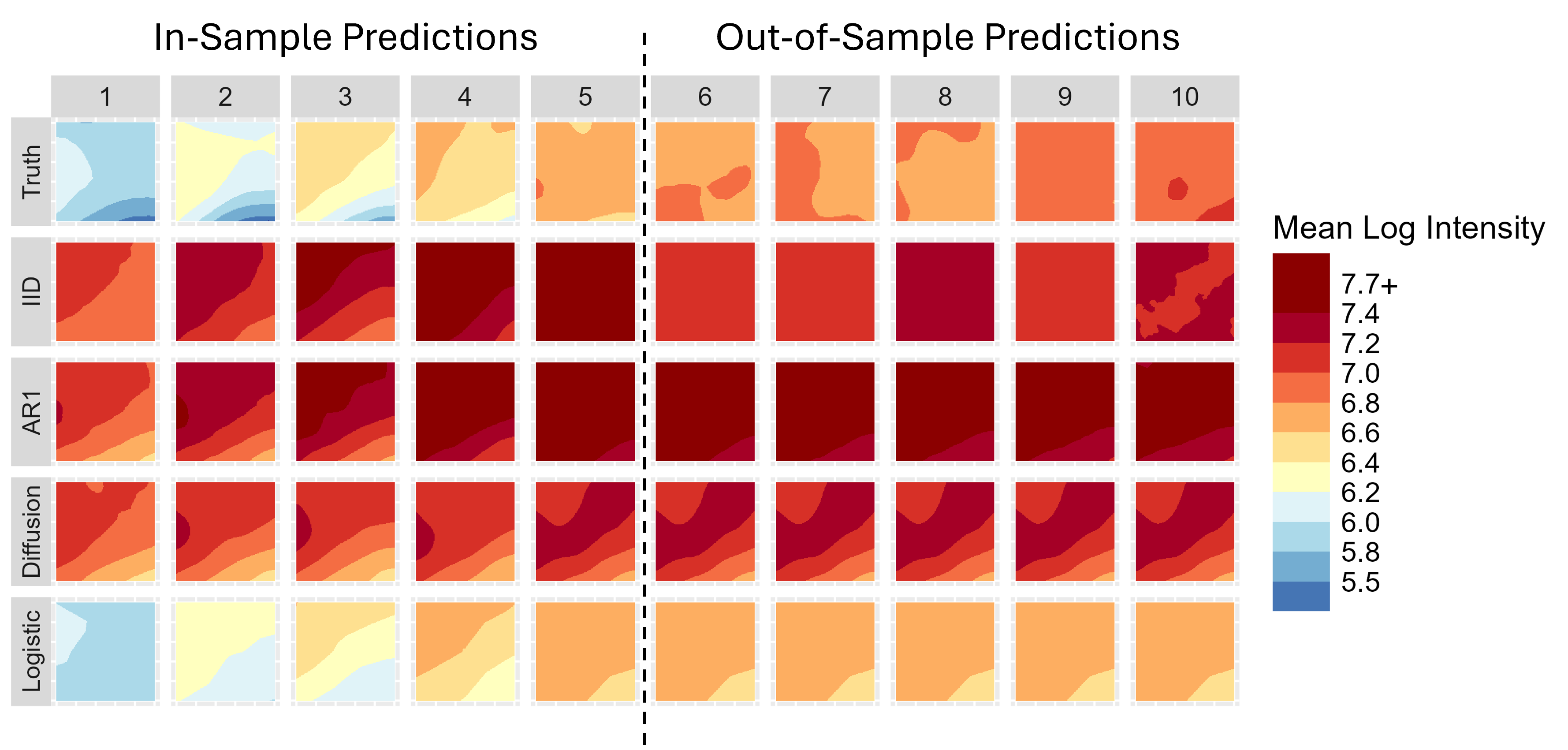}
    \caption{True abundance and mean predicted log intensity (abundance) for all models for all 10 years. All plots are on the same binned colour scale.}
    \label{fig:all-preds}
\end{figure}

The logistic growth model recovered accurate estimates for all parameters (Table \ref{tab:param-post}). All models returned credible intervals containing the true value for the growth and carrying capacity. Estimates of the precision parameter were more variable across models. The IID, diffusion, and logistic growth models all returned credible intervals that included the true value for the precision parameter, although uncertainty was substantially greater for the diffusion model. In contrast, the AR1 model yielded a narrower interval that excluded the true value. The separable models tended to slightly overestimate the carrying capacity and had large uncertainties for the growth rate, causing inaccurate in-sample and out-of-sample predictions that overestimated the abundance (Figure \ref{fig:all-preds}). The non-separable diffusion model performed slightly better as it was able to capture some of the complexity in the changing spatial pattern, but was not able to predict the dispersing behaviour in the out-of-sample predictions. 

\begin{table}[hbt!]
    \begin{tabular}{c|c|c|c|c}
         Parameter & Growth $\alpha$ & Carrying Capacity $K$ & Range/Dispersal $\xi$ & Precision $1/\sigma$ \\ \toprule
         Truth & 0.8 & 1000 & 0.15 & 5 \\
         IID model & 0.82 (0.33,1.31) & 1307 (886, 1698) & 2.75 (1.17, 5.28) & 3.24 (2.11, 5.38) \\
         AR1 Model & 0.78 (0.32, 1.25) & 1168 (805, 1524) & 2.53 (2.41, 2.66) & 2.86 (1.94, 4.42) \\
         Diffusion & 0.82 (0.548, 1.095) & 1088 (717, 1459) & 2 (0.62,7.39) & 16 (0.56, 469) \\
         Logistic Growth & 0.54 (0.30, 0.96) & 953 (693,1308) & 0.11 (0.05, 0.20) & 3.75 (2.19,6.45)
    \end{tabular}
    \caption{Posterior means and 95\% credible intervals for each model parameter}
    \label{tab:param-post}
\end{table}

\begin{figure}[hbt!]
    \centering
    \includegraphics[width=\linewidth]{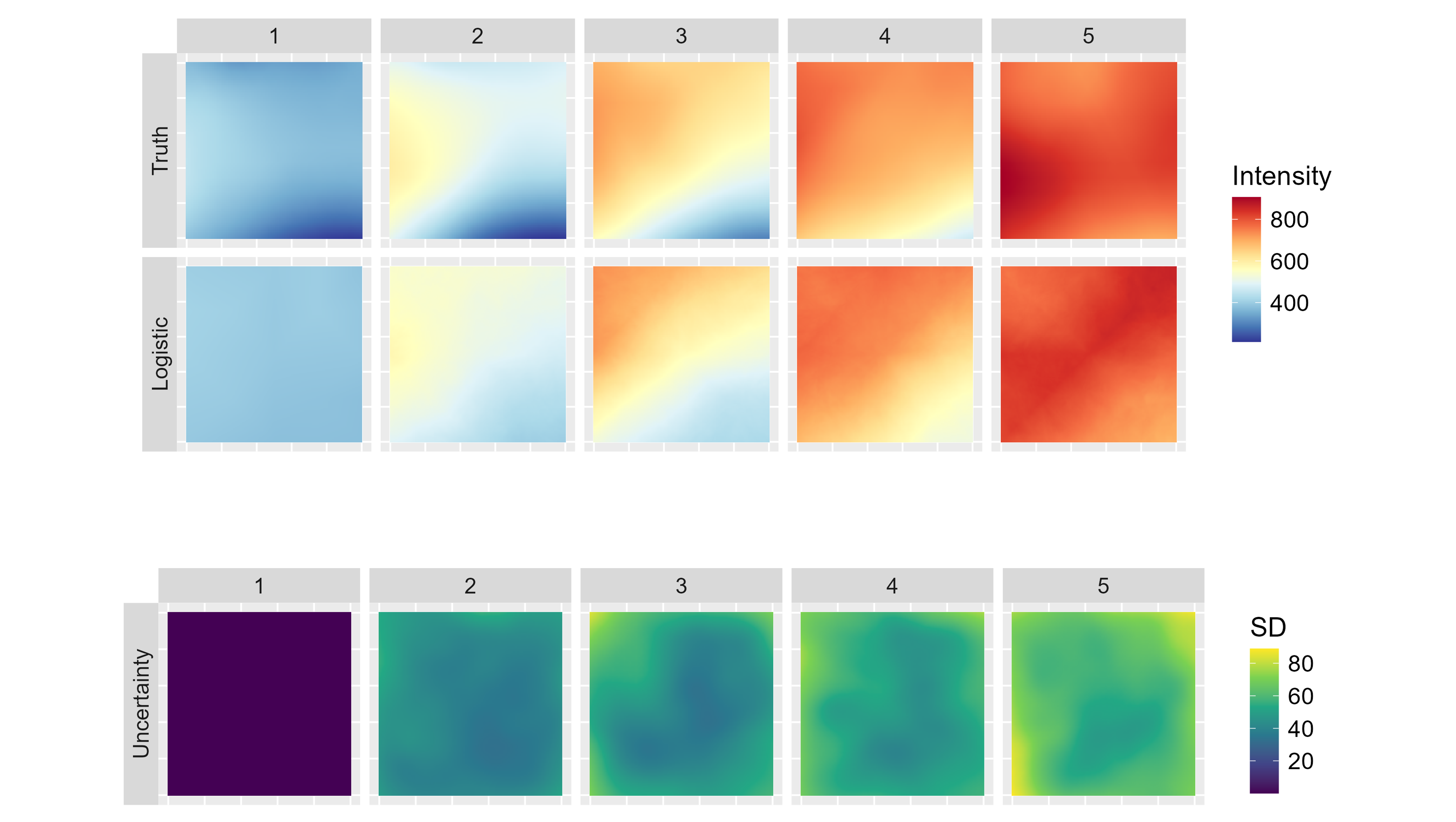}
    \caption{In sample predictions (years 1-5) for the logistic growth model. From top to bottom: mean predicted intensity, true intensity, and standard deviation of the predictions}
    \label{fig:loggrow-preds}
\end{figure}

\begin{figure}[hbt!]
    \centering
    \includegraphics[width=\linewidth]{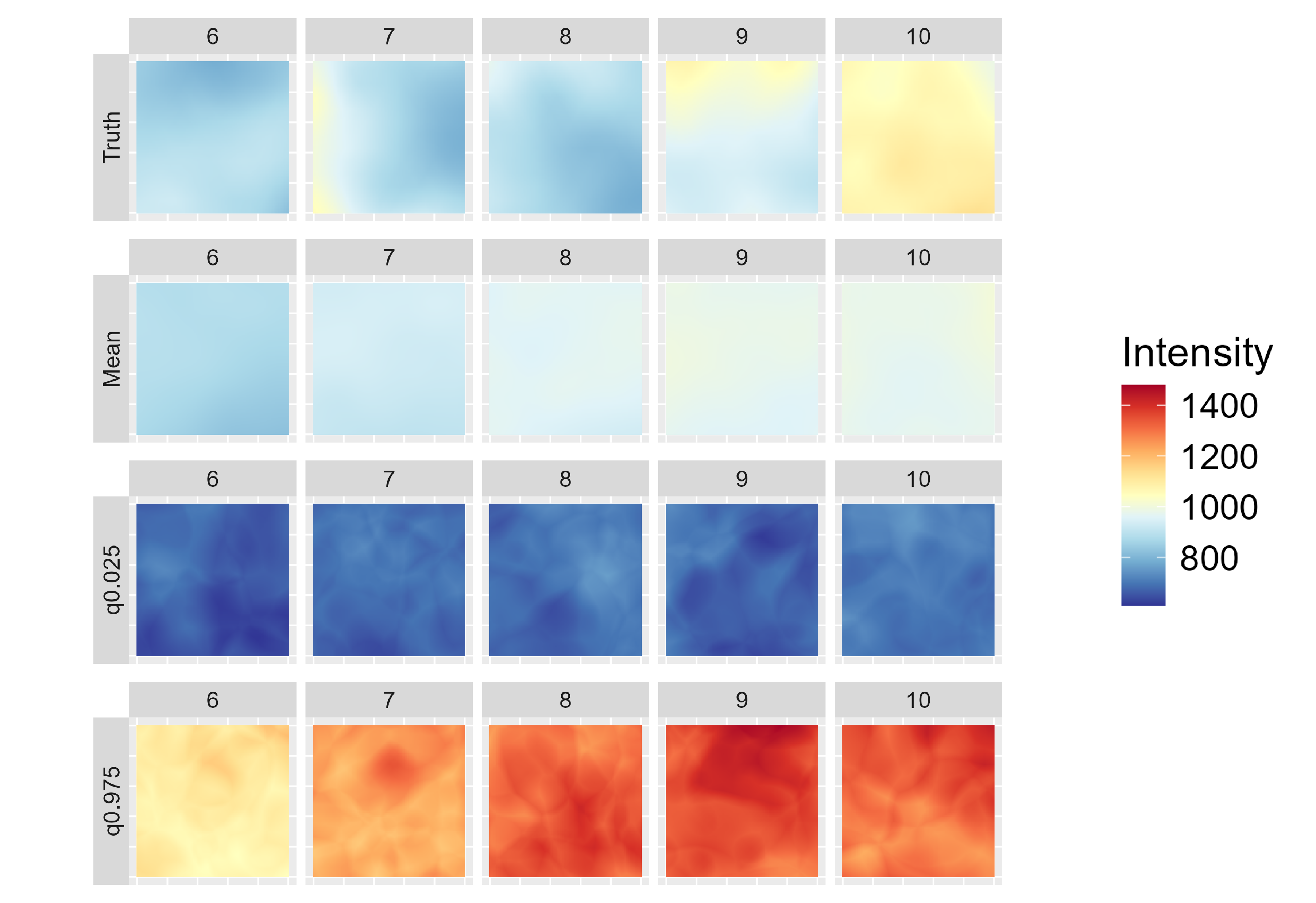}
    \caption{Out of sample forecasts (years 6-10) for the logistic growth model. 100 predictions were generated from the model. From top to bottom: the true intensity, the mean predicted intensity, the 2.5\% quantile predicted intensity and the 97.5\% quantile predicted intensity. All plots are shown on the same scale}
    \label{fig:preds-outofsample}
\end{figure}

The proposed logistic growth model gave much more accurate in-sample predictions (Figure \ref{fig:loggrow-preds}), capturing the abundance and distribution of the population. It also gave better out-of-sample forecasts than the other models, with the true intensity being inside the 95\% prediction interval for all years (Figure \ref{fig:preds-outofsample}). This is reflected in the mean SSIM value being closest to 1 (Table \ref{tab:gof-metrics}), showing that the logistic growth predictions are the most similar to the true intensity.

The logistic growth model outperforms the other models in all the posterior predictive scores (Table \ref{tab:gof-metrics}).  The two separable models in general have the highest posterior predictive scores, showing how separable models can struggle to capture spatial structures that evolve over time. The diffusion model performs better as it is able to capture some of the non-separable spatial-temporal dependence and diffusive behaviour, but not the non-linearity.
\begin{table}[ht]
    \centering
    \begin{tabular}{c|c|c|c|c}
         Metric & IID & AR1 & Diffusion & Logistic Growth \\ \toprule
         In-sample MSE& 1.01& 1.01 &0.60 & \bf{0.01} \\
         Out-of-sample MSE & 0.11& 0.53&0.16 & \bf{0.04} \\ \midrule
         In-sample MAE& 1.01& 1.00&0.76 &\bf{0.06}  \\
         Out-of-sample MAE & 0.35&0.712 &0.39 &\bf{0.17}  \\ \midrule
         In-sample CRPS& 43.7 &43.9 &29.6 &\bf{2.74} \\
         Out-of-sample CRPS& 9.32 & 26.5 & 12.2 &\bf{2.24}  \\ \midrule
         Mean SSIM & 0.778 & 0.699 & 0.797 & \bf{0.901}
    \end{tabular}
    \caption{Posterior predictive scores and mean SSIM for the 4 models. For the posterior predictive scores (MSE, MAE, CRPS), lower values indicate better predictive performance. SSIM ranges from -1 and 1, with higher values indicating greater similarity between the true map and the mean posterior map; a value of -1 represents complete dissimilarity while a value of 1 indicates identical maps. The best score for each metric is in bold. }
    \label{tab:gof-metrics}
\end{table}

\subsection{Case Study - Eurasian Collared Dove}\label{sec:dove}

To demonstrate the potential utility of this logistic growth model for invasive species, we apply the model to a case study of the Eurasian Collared Dove (\emph{Streptopelia decaocto}) in the United States (U.S.). We use data from the North American Breeding Bird Survey (BBS) between 1998 and 2003 \citep{BBS-2026}, previously used in logistic growth models by \cite{Hooten-2008} and \cite{Scheidt-2014}.
The dataset provides a particularly suitable case study because the species' introduction and range expansion in North America are well documented. First introduced to the Bahamas in 1974, this species since spread to Florida in the late 1970s and early 1980s and from there to the rest of the US \citep{Hengeveld-1993}.  The species underwent rapid population growth and spatial spread following its introduction, making it an example of the type of growing and dispersing population that the logistic growth equation is intended to describe. 
\begin{figure}
    \centering
    \includegraphics[width=\linewidth]{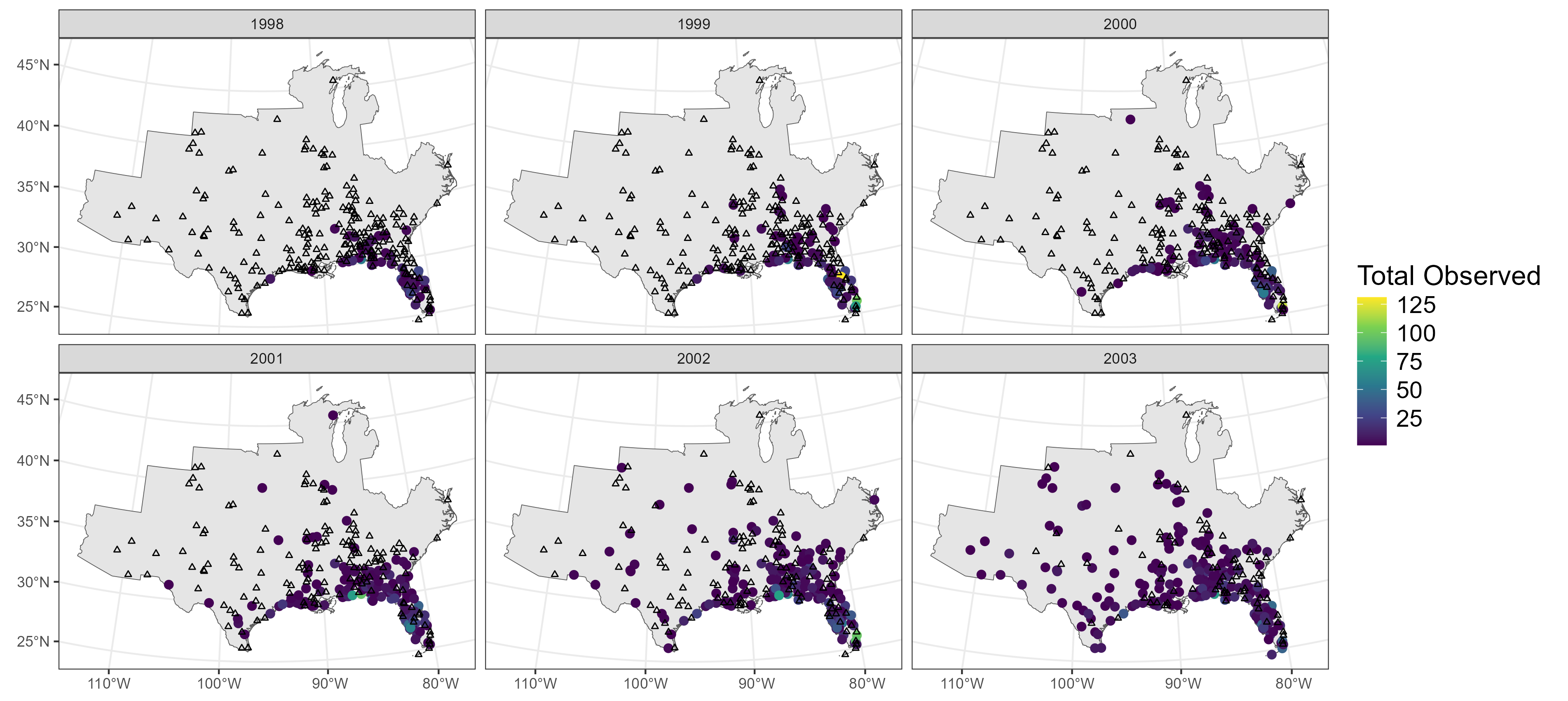}
    \caption{European Collared Dove counts, 1998-2003. Triangles are transects where no doves were observed; circles are transects where at least one dove was seen.}
    \label{fig:dove-data}
\end{figure}

The BBS data consist of yearly counts from approximately 39.2km long transects across North America. Observers stop approximately every 800m to conduct a 3-minute point count, repeated 50 times along the transect. We take the total number of Collared Doves observed on each transect (Figure \ref{fig:dove-data}). Following \cite{Hooten-2008} and  \cite{Scheidt-2014}, we discard all transects where Collared Doves were never seen in the study period, giving 269 transects where Collared Doves were seen at least once between 1998-2003. A large proportion (58\%) of the counts are zeros. Some of these are ``true'' zeros, where the doves were not observed because they were not present. Other zero counts reflect imperfect detection, where Eurasian Collared Doves were present but not seen by the recorders. 
To account for these excess zeros, we use a zero-inflated Poisson model, which allows zeros to arise from two separate processes: a zero-generating process and a Poisson count process. We take the log intensity of the Poisson part to be the proposed logistic growth model $v(\bm{s},t)$.

\begin{align}
    \mathbb{P}(y_{\bm{s},t}|\lambda, \bm{\theta}) = p\times 1_{y_{\bm{s},t}=0} + (1-&p)\times \frac{\exp{(-\lambda(\bm{s},t))}\lambda(\bm{s},t)^{y_{\bm{s},t}}}{y_{\bm{s},t}!}\\ \nonumber
    \log(\lambda(\bm{s},t)) &= v(\bm{s},t)
\end{align}

\subsubsection{Results}

The carrying capacity posterior (Table \ref{tab:dove-posteriors}) is broadly in line with the previous studies on these data. \cite{Scheidt-2014} estimated individual carrying capacities for each transect using logistic and quadratic growth models for the carrying capacities. They estimate carrying capacities of between 0 and 93 doves/transect, with most falling in the 0-30 range \citep[Fig.3]{Scheidt-2014}. Our model estimates a mean carrying capacity of 13 (12,14) doves/transect, which fits within this range. As our estimate is the mean carrying capacity across all sites, it captures less variability than the individual site estimates of \cite{Scheidt-2014}. It is difficult to directly compare the other parameters with the previous studies, as \cite{Scheidt-2014} estimate the individual transect growth rates using an exponential growth model, so obtain lower estimates of between 0 and 0.3. \cite{Hooten-2008} estimates the growth rate to be very close to 0 ($<0.02$ \cite[Fig. 3]{Hooten-2008}); this seems unlikely as the population is clearly expanding and is possibly due to the zero-inflated nature of the data. Our estimate of 1.65 (1.62, 1.69) suggests that the population is growing steadily, in keeping with existing literature \citep{Romogosa-2000, Hengeveld-1993}.

 \begin{table}[ht]
        \centering
        \begin{tabular}{c|c|c}
            Parameter & Posterior Mean & 95\% Credible interval \\ \toprule
            Growth Rate & 1.65 & 1.62, 1.69 \\
            Carrying Capacity & 13 & 12, 14 \\
            Dispersal (km) & 30.6 & 29.9, 31.4 \\
            Sigma & 2.6 & 2.5, 2.7 \\
            Zero-probability parameter & 0.28 & 0.27, 0.29\\
        \end{tabular}
        \caption{Parameter posteriors for the Eurasian Collared Dove data}
        \label{tab:dove-posteriors}
    \end{table}

\begin{figure}[hbt!]
    \centering
    \includegraphics[width=\linewidth]{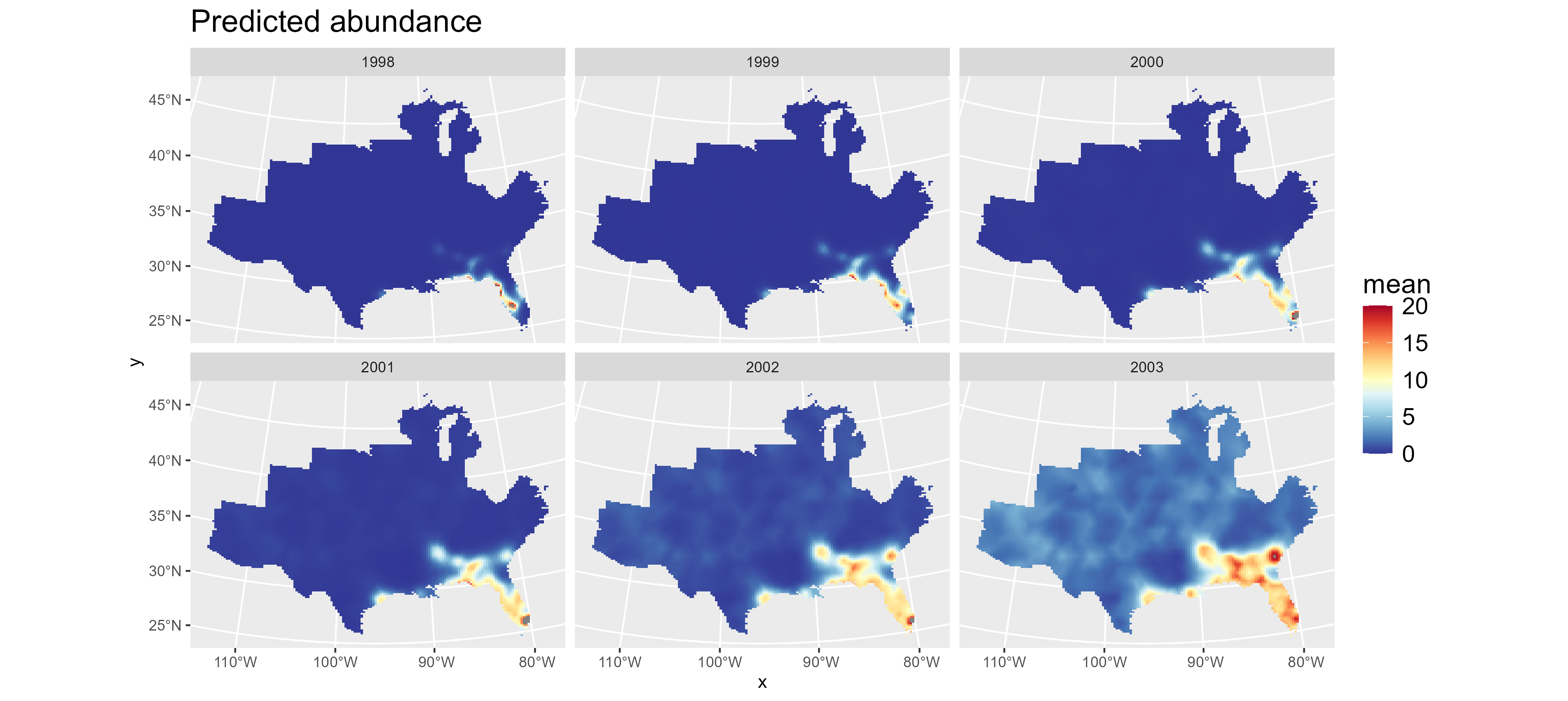}
    \caption{Mean predicted number of doves seen at a BBS transect 1998-2003}
    \label{fig:dove-abund}
\end{figure}

The predicted abundance (Figure \ref{fig:dove-abund}) shows the doves establishing a stable population in the South East and spreading to the North and West as time goes on, similar to the findings of \cite{Hooten-2008} and \cite{Scheidt-2014}. These results are consistent with the known rapid expansion of the Eurasian Collared Dove across North America and demonstrate that the proposed model can recover biologically plausible growth and dispersal dynamics from zero-inflated count data. This illustrates the potential value of combining mechanistic non-linear process models with the computational efficiency and flexibility of the INLA framework.

\begin{figure}[hbt!]
    \centering
    \includegraphics[width=\linewidth]{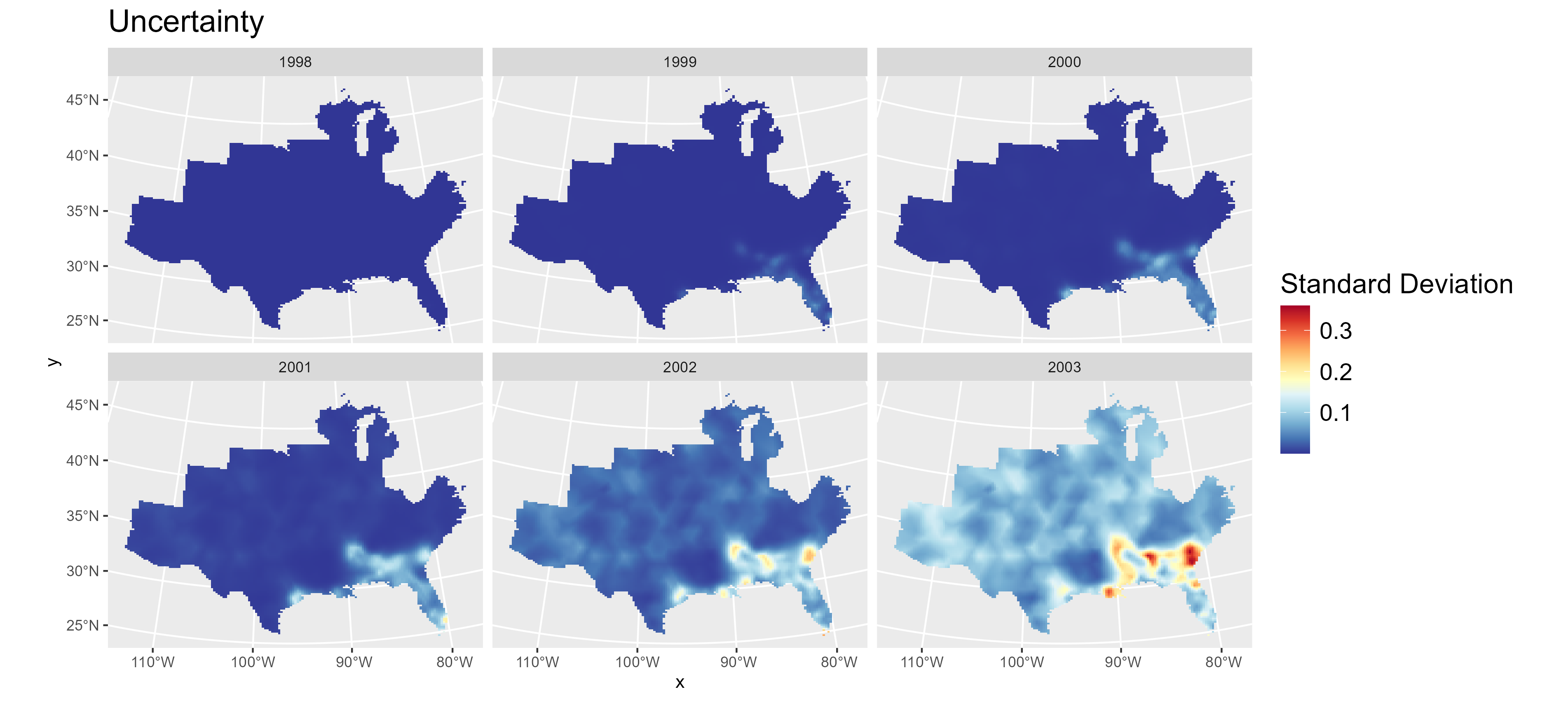}
    \caption{Standard deviation of the predicted number of doves seen at a BBS transect 1998-2003}
    \label{fig:placeholder}
\end{figure}

\section{Conclusion}\label{sec:conc}
In this study, we developed a framework for deriving a Gaussian Markov random field from a two-dimensional non-linear SPDE, enabling the development of mechanism-informed statistical models for a broad range of spatio-temporal applications. By incorporating knowledge of the underlying dynamics directly into the latent model, these process-informed statistical models enable inference on mechanistically meaningful and interpretable parameters that provide insight into system behaviour. They can also improve out-of-sample predictive performance relative to the purely statistical alternatives. We further extend previous methodology to accommodate non-Gaussian observation processes, substantially broadening the range of data types and domains to which the framework can be applied. 

\subsection{Logistic Growth}
We demonstrated the utility of this method with a specific application of the logistic growth equation, motivated by its use in ecological contexts. Our simulation study shows that incorporating non-linear dynamics into  spatio-temporal covariance functions substantially improves model performance relative to commonly used separable and diffusion‑based alternatives. The proposed model was the best across all predictive scores, showing that it could be a useful tool for predicting species distributions across time. As the proposed model was also the data-generating process, some improvement in predictive performance is expected; however, it consistently outperformed the competing models across all predictive scores. The case study of the Eurasian collared dove demonstrates that by extending the method to all non-Gaussian observation processes implemented in \texttt{R-INLA}, we are able to easily account for zero-inflation and therefore obtain more realistic parameter estimates than previous studies. The computational advantages of the finite element discretisation allow calculating predictions at a finer spatial resolution than previously possible. Recent work has shown that finite element approximations can be more accurate than physics-informed neural networks \citep{Grossmann-2024}, providing further support for the use of the SPDE approach in mechanism-informed modelling. 

One limitation of the logistic growth model is its sensitivity to model misspecification. \cite{Browing-2025} demonstrated that spatio-temporal logistic growth models can produce narrow but incorrect parameter posterior distributions when fitted to data generated by non-logistic processes. Hence, parameter estimates should be interpreted with caution. Future work could make this model more robust to misspecification, potentially by building on \cite{Browing-2025}'s work with Gaussian processes. 

\subsection{Complexity challenges}
Non-linear spatio-temporal models are inherently complex. The latent field needs to be estimated at each of the hundreds or thousands of mesh nodes at each point in time, in addition to estimating the model parameters. While structural identifiability issues can be addressed through careful specification of the underlying PDE, the quantity and quality of data required for practical identifiability may be a limitation in some applications. In the context of the logistic growth model, the Eurasian Collared Dove case study worked well due to the large number of observations and the structured nature of the surveys across the US. For species in the early stages of invasions with sparse records and/or noisier observation processes (e.g.\ some citizen science programs or instances where multiple data sources with different observation processes need to be integrated), identifiability issues may be more prevalent. 

The complexity of the model also presents scalability challenges. In principle, a finer temporal discretisation improves the accuracy of the backwards Euler approximation. However, each additional time point increases the size of the precision matrix by $n_s$ rows and columns, where $n_s$ is typically in the order of hundreds and thousands. As a result, repeatedly calculating and inverting the precision matrix becomes computationally demanding, with the cost increasing rapidly as the temporal resolution becomes finer. For example, fitting the model to the dove data with 1,396 mesh nodes and six time points required 70 minutes and 44 GB of memory on four CPUs. Consequently, although this framework can accommodate fine spatial and temporal discretisations, doing so in practice requires substantial computational time and resources. 

\subsection{Future Research}
There remain several promising avenues for future research. Extending the model to allow spatially varying hyperparameters by using non-stationary SPDE approaches \citep{Ingebrigtsen-2014, Mandujano-Reyes-2024} would allow for more realistic inference, and in the context of the logistic growth equation, would enable assessments of how environmental covariates affect population dynamics. Our methodology could be also used for other more complex non-linear SPDEs in different contexts. For example, the Allee effect \citep{Allee-1931} could be introduced into ecological models to incorporate the idea of a minimum sustainable population density, which is particularly relevant for the removal of invasive species or conservation of critically endangered populations. There is also potential outside of ecology; for instance, advection-diffusion equations could allow for more realistic models of pollutant dispersal \citep{Stockie-2011}. Our framework makes it possible for any 2D univariate non-linear differential equation to become the basis of a flexible statistical model, allowing statisticians to ground their models in the physical reality of the data generating process using the broad array of mechanistic modelling literature. However, this is not without challenges. In particular, calculating the precision matrix in a computationally efficient way is not trivial for the logistic growth model, and similar difficulties would likely be present for the non-stationary model or other non-linear SPDEs. Further work is therefore needed to investigate efficient and identifiable implementations for these more complex model classes.

\bigskip

\begin{center}
{\large\bf SUPPLEMENTARY MATERIAL}
\end{center}

\begin{description}

\item[Appendix:] Proof of Type II update rule for non-Gaussian likelihoods, and derivation of the logistic growth model.

\item[R package for spatial logistic growth model:] R-package INLAloggrowth containing code to fit the logistic growth model described in the article and the simulated data. Available at \url{https://github.com/ljb34/INLALogisticGrowth}

\end{description}

\bibliographystyle{apalike}

\bibliography{Bibliography-MM-MC}


\end{document}


\def\spacingset#1{\renewcommand{\baselinestretch}%
{#1}\small\normalsize} \spacingset{1}

\title{\bf Supplementary Material}
\author{L.\ Blackman$^{1,2}$\thanks{LB, PB and EJ were supported by the Scottish Government's Rural and Environment Science and Analytical Services Division (RESAS)}\hspace{.2cm}, J.\ Belmont$^1$, P.J.\ Bouchet$^{2*}$,J.\ Illian$^1$, E.L.\ Jones$^{2*}$,  F.\ Lindgren$^3$  \\ \small{$^1$ School of Mathematics and Statistics, University of Glasgow, UK}
     \\     
     \small{$^2$ Biomathematics and Statistics Scotland, UK}\\ 
     \small{$^3$ Lund University}
}
  \maketitle

\section{Type II update for Point Processes and other observation processes}
The Type II update rule uses the Gauss Newton Method to find the linearisation point $u_0$ that minimises $\mathbb{E}(-\log p(\bm{u|y,\theta}))$. \cite{Anderka-2024} proves this for one-dimensional Gaussian observations, here we shall prove that it can be extended to any observation process that can be adequately approximated using the Laplace approximation. We focus on Log Gaussian Point Process observations both because our example use case is a point process, and because it has a more involved approximation than the other observation families in \texttt{inlabru}, so by proving the LGCP case the others follow naturally. Let $\bm{y}$ be observations from a Log Gaussian Cox Process (LGCP) with log intensity $\bm{u}$ governed by a non-linear SPDE model: 
\begin{align}
    \bm{y|u, \theta} &\sim \text{Poisson Process}(\exp(\bm{u})) \\
    \bm{u|\theta} &\sim N(\bm{L}^{-1}\bm{r}, (\bm{L}^\top\bm{Q}^{-1}\bm{L})^{-1}) \\
    \bm{\theta} &\sim p(\bm{\theta})
\end{align}
$\bm{L,Q}$ and $\bm{r}$ depend on the linearisation point $u_0$, which we wish to find iteratively using the Type II update rule. By the Gauss-Newton Method:
\begin{equation}
    u_0^{(n+1)} = u_0^{(n)} - \gamma \bm{B}^{-1} \nabla_u \mathbb{E}(-\log p(\bm{u|y,\theta}))
\end{equation}
where B is the approximate Hessian of $p(\bm{u|y,\theta})$. For LGCP observations $\bm{y}$: 
\begin{equation}
    \log p(\bm{u|y,\theta}) = -\frac{1}{2} \mathcal{L}^\top(\bm{u})Q^{-1}\mathcal{L}(\bm{u}) -a^\top \exp{A_1\bm{u}} + 1^\top A_2\bm{u} + C
\end{equation}
Using the $\mathcal{L}(u)$ notation from \cite{Anderka-2024}, $\mathcal{L}(u) = L(u-L^{-1}r)$ and the \cite{Simpson-2016} formulation of the LGCP likelihood. $a$ is the vector of integration weights, $A_1$ is the mapper matrix to the mesh nodes, and $A_2$ the mapper matrix to the observed locations. $\bm{u}$ is only defined on the mesh nodes, so $A_1 \equiv I$.
\begin{align}
    \nabla_u \mathbb{E}(-\log p(\bm{u|y,\theta})) &= \mathbb{E}((\nabla\mathcal{L}(\bm{u}))^\top Q^{-1}\mathcal{L}(\bm{u}) + a \odot \exp{\bm{u}} - A_2^\top 1) \\
    \nabla^2_u \mathbb{E}(-\log p(\bm{u|y,\theta})) &= \mathbb{E}((\nabla\mathcal{L}(\bm{u}))^\top Q^{-1}(\nabla\mathcal{L}(\bm{u})) + \diag(a \odot \exp{\bm{u}})) := \bm{B}
\end{align}

where $\odot$ denotes pairwise multiplication, $a \odot \exp{\bm{u}} = (a_1\exp{u_1}, a_2\exp{u_2},...,a_n\exp{u_n} )^\top$. 

\begin{align}
    u_0^{(n+1)} &= u_0^{(n)} - \gamma\bm{B}^{-1} \mathbb{E}((\nabla\mathcal{L}(\bm{u}))^\top Q^{-1}\mathcal{L}(\bm{u}) + a \odot \exp{\bm{u}} - A_2^\top 1 +\bm{B}u_0^{(n)} - \bm{B}u_0^{(n)}) \\
    &= (1-\gamma)u_0^{(n)} - \gamma \bm{B}^{-1}\mathbb{E}((\nabla\mathcal{L}(\bm{u}))^\top Q^{-1}(\mathcal{L}(\bm{u})-\nabla\mathcal{L}(\bm{u})u_0^{(n)}) + a \odot \exp{\bm{u}} \\ &- \diag(a \odot \exp{\bm{u}})u_0^{(n)} - A_2^\top1) 
 \end{align}

 Substituting $\nabla\mathcal{L}(\bm{u}) = L$
 \begin{align}\label{eq:A1-update-long}
     u_0^{(n+1)} = (1-\gamma)u_0^{(n)} + \gamma \bm{B}^{-1}\mathbb{E}(L^\top Q^{-1} L L^{-1}r + \diag(a \odot \exp{\bm{u}})u_0^{(n)} - a \odot \exp{\bm{u}} + A_2^\top 1)
 \end{align}

 Using the Laplace approximation, $\bm{u|y,\theta} {\sim} N(\bm{\mu},(L^\top Q^{-1}L + \diag(a \odot \exp{\bm{\mu}}))^{-1})$ approximately, where $\bm{\mu} = \argmax_u \log(p(\bm{u|y,\theta})$. Hence $\bm{\mu}$ satisfies $\nabla_u \mathbb{E}(-\log p(\bm{\mu|y,\theta})) = 0$ :
 \begin{align}
     L^\top Q^{-1}L(\bm{\mu} - L^{-1}r) + a \odot \exp{\bm{u}} - A_2^\top 1 = 0 \\
      L^\top Q^{-1}L\bm{\mu} = L^\top Q^{-1}LL^{-1}r + A_2^\top1 - a \odot \exp{\bm{u}} 
\end{align}
So we can simplify the expectation term in \eqref{eq:A1-update-long}
\begin{align} \label{eq:A1-simplification}
     \mathbb{E}(L^\top Q^{-1} L L^{-1}r &+ \diag(a \odot \exp{\bm{u}})u_0^{(n)} - a \odot \exp{\bm{u}} + A_2^\top 1) \\ \nonumber &= \mathbb{E}((L^\top Q^{-1}L + \diag(a\odot \exp{\bm{u}}))\bm{\mu}) \\ \nonumber
     &\approx \sum_k\bm{P_{u|y}}^{(n)}(\bm{\theta}_k^{(n)}) \bm{\mu_{u|y}}(\bm{\theta}_k^{(n)}) \tilde p(\bm{\theta_k}^{(n)}|\bm{y})\Delta_k \\ \nonumber
     &:= \bm{\bar{b}^{(n)}}
 \end{align}
As $\bm{B} \approx \sum_k \bm{P_{u|y}}^{(n)}(\bm{\theta}_k^{(n)})\tilde p(\bm{\theta_k}^{(n)}|\bm{y})\Delta_k := \bm{\bar{P}^{(n)}}$, the Gauss-Newton Method for minimising the log posterior is equivalent to the type II update rule: 
\begin{equation}
    u_0^{(n+1)} = (1-\gamma)u_0^{(n)} + \gamma(\bm{\bar{P}^{(n)}})^{-1}\bm{\bar{b}}^{(n)}
\end{equation}

This result will hold for other observation processes that can be adequately approximated using the Laplace approximation, not just point processes, because the simplification in \eqref{eq:A1-simplification} applies whenever the Laplace approximation is valid.]

\section{Derivation of the Logistic Growth Gaussian Random Field}
We take the following form of the logistic growth equation as the basis for our model, as described in Section 3:
\begin{equation} \label{eq:A2-growth-diffusion}
    \frac{\partial u}{\partial t} - \alpha u\left(1-\frac{u}{K}\right) + \nabla\cdot \xi \nabla u = \sigma uW(\bm{s},t).
\end{equation}

We take a log transformation $u = \exp(v)$ to enforce $u \geq 0$. Rewriting equation \eqref{eq:A2-growth-diffusion} in terms of $v$ and dividing through by $\exp(v)$, we gain an equation in the form of (10):
\begin{equation}\label{eq:A2-log-transf}
    \frac{\partial v}{\partial t} - \alpha \left(1-\frac{\exp(v)}{K}\right) + \xi \nabla^2 v + \xi ||\nabla v||^2 = \sigma W(\bm{s},t)
\end{equation}

We can then linearise \eqref{eq:A2-log-transf} using a first-order Taylor expansion about a linearisation point $v_0^{(n)}$:
\begin{equation}\label{eq:A2-linearised}
    \frac{\partial v}{\partial t} - \alpha  \left(1-\frac{\exp(v_0^{(n)})}{K}v \right) - \alpha \frac{\exp(v_0^{(n)})}{K}(v_0^{(n)}-1) + \xi \nabla^2v + \xi||\widehat{\nabla v_0^{(n)}}||^2 = \sigma W(\bm{s},t),
\end{equation}
where $\widehat{\nabla v_0^{(n)}}$ is a numerical approximation of $\nabla v$ at the linearisation point. 

Equation \ref{eq:A2-linearised}  can be rearranged into the same form as the diffusion injected Matérn equation in \cite{Lindgren-2024-SPDE}, but with an additional drift term $r_0^{(n)}$:  
\begin{equation}
    \left( \frac{\partial}{\partial t} - (a(v_0^{(n)})-\xi\nabla^2)\right) v(\bm{s},t)  = r_0^{(n)} +  \sigma W(\bm{s},t),
\end{equation}
where $a(v_0^{(n)}) = \alpha \exp(v_0^{(n)})/ K$ and $r_0^{(n)} = \alpha +  \alpha \exp(v_0^{(n)})(v_0^{(n)}-1)/K - \xi||\widehat{\nabla v_0^{(n)}}||^2$. This representation motivates choosing $W(\bm{s},t)$ to be Matérn correlated in space and white in time, as it places the model in the critical diffusion case, with smoothness $1/2$ in time and $1$ in space \citep{Lindgren-2024-SPDE}. Completely unstructured noise (i.e.\ white in space and time) would give smoothness 0 in both space and time, invalidating the assumptions of the numerical discretisation methods and generating an non-smooth field.  

Now, apply the finite element method discretisation in space:
\begin{equation}
    \left( \bm{C}\frac{\partial}{\partial t} + \bm{C}a(\bm{v}_0^{(n)}) + \xi\bm{G} \right) \bm{v}(t) = \bm{Cr}_0^{(n)}(t) +  \sigma \bm{CW}(t)
\end{equation}
where $C_{i,j} = \innerprod{\phi_i}{\phi_j}$ and $G_{i,j} = \innerprod{\nabla \phi_i}{\nabla \phi_j}$. In practice, to ensure a sparse representation and computational efficiency, $\bm{C}$ is approximated by the diagonal matrix $\tilde C_{i,i} = \innerprod{\phi_i}{1}$. Multiply through by $\bm{C}^{-1}$:
\begin{equation}
    \left( \frac{\partial}{\partial t} + a(\bm{v}_0^{(n)}) + \xi\bm{C^{-1}G} \right) \bm{v}(t) =  \bm{r}_0^{(n)}(t) + \sigma \bm{W}(t)
\end{equation}

Discretising in time using the backward Euler method and assuming $\bm{v}(T_1) \sim N(\bm{\mu}, \bm{P}^{-1})$ gives:

\begin{align} \label{eq:A2-matrix-form}
    &\left( \frac{1}{h}\begin{bmatrix}
        h\bm{I} & 0&\cdots & 0\\
        -\bm{I} & \bm{I} & &\vdots \\
        \vdots&\ddots & \ddots & \vdots \\
         0& \cdots&-\bm{I} & \bm{I}
    \end{bmatrix} + \begin{bmatrix}
        0 & \cdots&\cdots &  0 \\
        0 &a(\bm{v}_0^{(n)}(T_2)) & & \vdots \\
        \vdots& & \ddots &\vdots\\
         0 & \cdots& \cdots& a(\bm{v}_0^{(n)}(T_{n_t}))
    \end{bmatrix} \right. \\ \nonumber
    &+ \left. \begin{bmatrix}
        0 & \cdots & \cdots & 0 \\
        0 & \xi\bm{C^{-1}G} & & \vdots \\
        \vdots & & \ddots & \vdots \\
        0 & \cdots &\cdots &\xi\bm{C^{-1}G}
    \end{bmatrix}
    \right)\begin{bmatrix}
        \bm{v}(T_1) \\
        \vdots\\
        \vdots\\
        \bm{v}(T_{n_t})
    \end{bmatrix} \\ \nonumber
    &= \begin{bmatrix}
        \bm{\mu} \\
        \bm{r}_0^{(n)}(T_2) \\
        \vdots \\
        \bm{r}_0^{(n)}(T_{n_t})
    \end{bmatrix} + \begin{bmatrix}
        \bm{W_P} \\
        \bm{W}(T_2) \\
        \vdots \\
        \bm{W}(T_{n_t})
    \end{bmatrix}
\end{align}

From this, we define the discretised linear operator matrix $\bm{L}$, 
\begin{equation}
\bm{L} = \begin{bmatrix}
    \bm{I} & 0&\cdots & 0\\
        -\frac{1}{h}\bm{I} & \frac{1}{h}\bm{I}+a(\bm{v}_0^{(n)}(T_2))+\xi\bm{C^{-1}G} & &\vdots \\
        \vdots&\qquad \ddots & \ddots & \vdots \\
         0& \cdots&-\frac{1}{h}\bm{I}  \qquad & \frac{1}{h}\bm{I}+a(\bm{v}_0^{(n)}(T_{n_t}))+\xi\bm{C^{-1}G}
\end{bmatrix}
\end{equation}
the remainder vector $\bm{r_0^{(n)}}$,
\begin{equation}
    \bm{r_0^{(n)}} = \begin{bmatrix}
        \bm{\mu} \\
        \bm{r}_0^{(n)}(T_2) \\
        \vdots \\
        \bm{r}_0^{(n)}(T_{n_t})
    \end{bmatrix}
\end{equation}
and the precision matrix of the noise process
\begin{equation}
    \bm{Q} = \begin{bmatrix}
        \bm{P} & \bm{0} & \cdots & \bm{0} \\
        \bm{0} & \frac{\sigma^2(\bm{C} + g\bm{G})}{h} & &\vdots \\
        \vdots & & \ddots & \vdots \\
        \bm{0} & \cdots & & \frac{\sigma^2(\bm{C} + g\bm{G})}{h}
    \end{bmatrix}
\end{equation}
where $g$ is a spatial scale parameter for the Matérn correlated noise process. The choice of $g$ is somewhat arbitrary, so we take $g = \xi$ to link the range to the dispersal of the animals.

Then, from \eqref{eq:A2-matrix-form}, $\bm{v} \sim N(\bm{L^{-1}}\bm{r}_0^{(n)}, (\bm{L}^\top\bm{Q}\bm{L})^{-1})$.
\bibliographystyle{apalike}

\bibliography{Bibliography-MM-MC}